\documentclass[a4paper, twocolumn, 10pt, superscriptaddress, pra, aps]{revtex4-1}
\usepackage[T1]{fontenc}
\usepackage[utf8]{inputenc}
\usepackage{lmodern}

\usepackage{multirow}

\usepackage{graphicx} % Allows including images
\usepackage{amsmath}
\usepackage{amsfonts} % para usar identidad=\mathbb{I}\sqrt{}
\usepackage{amssymb}  % assumes amsmath package installed
\usepackage{mathtools}
\usepackage{booktabs}

\usepackage{natbib}
\usepackage{color}   %May be necessary if you want to color links
\usepackage{xcolor}
\definecolor{celeste}{cmyk}{1.00, 0.00, 0.00, 0.4}
\usepackage{hyperref}
\hypersetup{
	colorlinks=true, %set true if you want colored links
	linktoc=all,     %set to all if you want both sections and subsections linked
	breaklinks=true, %allow line break inside links
  linkcolor=celeste, %   - del table of contents, y los links a las figuras y ecuaciones.
  citecolor=celeste,  %   - de las citas en el texto 	
  urlcolor=celeste
}

\newenvironment{lcase}
{\left\lbrace \begin{aligned}}
	{\end{aligned} \right.}
\DeclarePairedDelimiter\abs{\lvert}{\rvert}%
\DeclarePairedDelimiter\expec{\langle}{\rangle}%
\begin{document}

\title{Entanglement across sliding-pinned transition of ion chains in optical cavities}

\author{Alan Kahan}
\author{Cecilia Cormick}
\affiliation{Instituto de F\'{i}sica Enrique Gaviola, CONICET and Universidad Nacional de C\'{o}rdoba,
Ciudad Universitaria, X5016LAE, C\'{o}rdoba, Argentina}

\date{\today}

\begin{abstract}
Dissipative quantum systems can under appropriate conditions exhibit bi- or multi-partite entanglement at the steady state. The presence and properties of these quantum correlations depend on the relevant model parameters. Here, we characterize the steady-state entanglement in connection with the spatial structure of a small chain of three ions dispersively coupled with a pumped optical cavity. %In this system, the optical potential can give rise to dynamics similar to that of the Frenkel-Kontorova model, where a transition from a sliding to a pinned phase is observable. 
Within a semiclassical approximation, we describe the relation between entanglement, spatial organization, and vibrational modes of the ion chain. Upon increasing the pumping strength, our system undergoes a transition from a sliding to a pinned configuration, in which ions are expelled from the maxima of the optical potential. The features of the steady-state entanglement strongly depend on the kind of pinned configuration reached. We identify scenarios leading to entangled steady states, analyze the effect of defect formation upon entanglement between different system partitions, and observe the presence of multipartite quantum correlations.
\end{abstract}

\maketitle

%%%%%%%%%%%%%%%%%%%%%%%%%%%%%%%%%%%%%%%%%%%%%%%%%%%%%%%%%%%%%%%%%%%%%%%%%%%%%%%%%%%
\section{Introduction}

The exceptional degree of controllability of trapped ions makes them ideal candidates for quantum simulations of condensed matter models \cite{Altman_2021, Porras_2004, Monroe_2021, Bermudez_2013, Mezzacapo_2012}.
In particular, the optomechanical interaction of ion chains and standing optical fields makes these systems particularly well-suited for the study of nano-friction \cite{Garcia-Mata2007, Benassi2011,gangloff2015velocity,bylinskii2015, bylinskii2016observation,kiethe2017,kiethe2018,timm2021, bonetti2021, chelpanova2024,menu2024}. 

Indeed, the optical field can induce in one-dimensional chains transitions   % \cite{Schmied_2008,cormick2012structural} 
that are closely related to the Frenkel-Kontorova (FK) model \cite{braun2013}.
In this model, a chain of masses connected by springs sits on a periodic potential.
The competition between the interparticle interaction and the substrate potential can give rise to a variety of ground-state configurations that depend both on the boundary conditions and on the ratio between the period of the substrate and the natural length of the springs. 
Possible variations of the FK model with trapped ions involve periodic substrates given by the potential of a pumped optical cavity \cite{gangloff2015velocity,bylinskii2015, bylinskii2016observation} or optical lattice \cite{chelpanova2024,menu2024}, or with a deformable potential provided by a second, neighboring ion chain \cite{kiethe2017,kiethe2018,timm2021}.

\begin{figure}[htpb!]
\centering
\includegraphics[width=1\columnwidth]{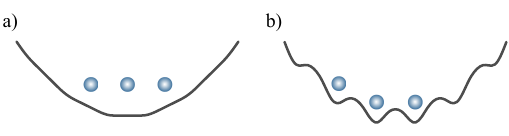}
\caption{Sketch of the sliding-pinned transition with three ions in a harmonic trap combined with an optical potential. In a) the optical potential is weak and equilibrium positions are mostly determined by the combination of harmonic confinement and electrostatic repulsion. b) When the optical potential becomes dominant, the ions are pushed towards optical minima.}
\label{esquema_transicion_FK}
\end{figure}

When the substrate potential is generated by the interaction with an optical cavity, the transition can be driven by increasing the laser pumping strength, as schematized in Fig. \ref{esquema_transicion_FK}.
Loosely speaking, the system is said to be in a sliding phase when the ions can be found at arbitrary locations relative to the optical field. On the contrary, for strong laser pumping the ions tend to localize at minima of the optical potential, which is usually referred to as the pinned phase. Due to the strong Coulomb repulsion, typical inter-ion distances are in practice of the order of dozens of lattice sites.

Several studies of friction and sliding-pinned transitions in trapped-ion platforms have been carried out along these lines, both theoretically and experimentally, analyzing many features such as equilibrium configurations, energy barriers, final temperatures, etc. \cite{Garcia-Mata2007, Benassi2011,gangloff2015velocity,bylinskii2015, bylinskii2016observation,kiethe2017,kiethe2018,timm2021, bonetti2021, chelpanova2024,menu2024,fogarty2015, Fogarty_2016}. Notably, the quantum correlations present in these setups remain to our knowledge largely unexplored. In this work we take a step in this direction, addressing the question of how the changes in spatial organization impact on the entanglement within the system. 

More precisely, we focus on steady-state entanglement in a one-dimensional chain of three trapped ions coupled with an optical cavity, across the transition from a sliding to a pinned configuration. Even though the problem we study is not in the many-body realm, it represents a minimal implementation of the more complex many-particle system. We characterize the continuous-variable (CV) entanglement 
\cite{adesso2006,lami2018} between various degrees of freedom and relate it with the equilibrium structure. The features of multipartite CV entanglement not only bring to light another aspect of the system dynamics; they could also be relevant in connection with the continuous-variable paradigm of quantum computation \cite{gottesman2001, Menicucci_2006}. Entanglement for ultracold neutral atoms in cavities has been studied in \cite{Maschler_2007, Kresic_2023}; systems of trapped ions are very different because of their strong Coulomb interaction.

An important precedent of this work is the study in~\cite{cormick2013}, which estimated the CV entanglement in the transition from linear to zigzag structures of ions in optical cavities. This analysis found non-negligible entanglement, although in parameter regimes for which the underlying semiclassical approximations could be unreliable \cite{Kahan_2021, kahan2023}. Besides, only bipartite entanglement between cavity and motional modes was studied. For the small system we consider here, we show that it is possible to obtain entangled steady states far from the regime where the semiclassical description is expected to fail. We also go beyond previous articles \cite{cormick2013, fogarty2015, Fogarty_2016} by analyzing both bi- and multi-partite entanglement for different system partitions.

We observe that the presence of defects in the chain structure has a direct correlation with the vibrational modes exhibiting most entanglement with the cavity field fluctuations. Furthermore, we find genuine tripartite entanglement within the ion chain, as well as four-mode entanglement when considering partitions involving the cavity and the local degrees of freedom of individual ions.

This work is organized as follows: in Sec.~\ref{sec:model} we describe the model and the equations governing the dynamics.  In Sec.~\ref{sec:semiclassical} we review the semiclassical treatment of the system. In Sec.~\ref{sec:sp_transition} we identify the different semiclassical equilibrium configurations as a function of the system parameters. Section~\ref{sec:entanglement} contains the main results, characterizing bi- and multi-partite CV entanglement between several system partitions. Finally, in Sec.~\ref{sec:conclusions} we summarize our results and discuss possibilities for future work. Additional details are provided in two appendices.

%%%%%%%%%%%%%%%%%%%%%%%%%%%%%%%%%%%%%%%%%%%%%%

\section{Ion chain dispersively coupled with an optical cavity}
\label{sec:model}

The system we consider is similar to the ones studied in \cite{linnet2012, cetina2013, fogarty2015, gangloff2015velocity, Fogarty_2016, bylinskii2015,bylinskii2016observation,counts2017,laupretre2019}.
We focus on a small linear chain of three equal ions, with mass $m$ and charge $q$, inside an optical cavity with a single relevant mode, of frequency $\omega_c$, that is pumped by a laser. The propagation direction of this mode is assumed to coincide with the longitudinal axis of the ion chain.

We start by describing the cavity dynamics in absence of coupling to the ions. The unitary evolution of the cavity field in the frame rotating with the laser is determined by the Hamiltonian:
\begin{equation}
  H_{\textrm{cav}}=-\hbar \Delta_c a^\dagger a + i \hbar \eta \left(a^\dagger -a  \right)\,.
\end{equation}
\noindent The operators $a^\dagger$ and $a$ are the creation and annihilation operators of this field, respectively, while $\Delta_c=\omega_l - \omega_c$ is the detuning between the laser frequency and the cavity mode. 
Finally, $\eta$ is the pump strength, proportional to the amplitude of the laser; with no loss of generality we have assumed $\eta \in \mathbb{R}$.

The imperfect reflectivity of the mirrors forming the cavity leads to photon loss. 
This decay of the photon number at a rate of $2\kappa$ can be described by the superoperator
\begin{equation}
  \mathcal{D}_\kappa(\rho) = 2\kappa \left(a \rho a^\dagger - \frac{1}{2} \{a^\dagger a, \rho\} \right)
  \label{eq:photon losses}
\end{equation}
where we consider that no thermal excitations enter the cavity from the environment. 

When the ions form a linear chain along the $x$ axis, the Hamiltonian describing the axial motional degrees of freedom in the absence of the cavity is:
\begin{equation}\label{ec_Hiones}
  H_{\textrm{ion}} =  \sum_{j=1}^N \Big(\frac{p_j^2}{2m} + \frac{m\omega^2 x_j^2}{2} \Big)
    + \sum_{j=1}^N \sum_{k>j}^N \frac{q^2}{4\pi \epsilon_0}\frac{1}{|x_j - x_k|}\,.
\end{equation}
Here $p_j$ is the momentum of ion $j$ along the $x$ axis. The trap potential is considered in harmonic approximation, with $\omega$ the trap frequency, and the particles interact through Coulomb repulsion. 
Such a linear ion chain can be obtained when the trap confinement along the $y$ and $z$ axes is much tighter than along $x$. Furthermore, for linear configurations the coupling between the motion in axial and transverse directions can usually be neglected~\cite{Marquet_2003}.

We will model the effect of noise on the ion motion through a coupling with an environment acting directly upon the normal modes of the ion chain. This causes damping as well as heating \cite{leibfried2003}, and is described by
\begin{equation}\label{disipador_ion}
\begin{aligned}
\mathcal{L}_{\Gamma_i} =  \frac{\Gamma_i}{2} \Big[ & (N_i+1)\left(2 b_i \rho b_i^{\dagger}-\left\{b_i^{\dagger} b_i, \rho\right\}\right)  \\ 
          & + N_i \left(2 b_i^{\dagger} \rho b_i-\left\{b_i b_i^{\dagger}, \rho\right\}\right) \Big]  
\end{aligned}
\end{equation}

\noindent for the $i$-th vibrational mode, with ladder operators $b_i$, $b_i^\dagger$ and $i=1,2,3$.
We note that in presence of interaction with the cavity, the vibrational modes described by $b_i$, $b_i^\dagger$ are defined accordingly, as in \cite{cormick2013}.
For simplicity, we consider that noise on the ions acts uniformly on each vibrational mode, $\Gamma_i=\Gamma$ and $N_i=N$. Although this assumption is not physically realistic, it is convenient for our purposes since motional noise is not the focus of our work.

We now turn to the dispersive interaction between cavity and ions. The frequency $\omega_{c}$ of the cavity is assumed to be close to resonance with the frequency of a dipole transition between two atomic levels, but far enough from resonance to avoid populating the excited states. In this regime, the excited electronic states can be adiabatically eliminated \cite{molmer2007,marzoli1994}, so that our model describes an effective optomechanical system. More precisely, labeling $\Delta_0$ the detuning between laser and atoms, we assume  
$\abs{\Delta_0 } \gg \kappa, \gamma, \abs{\Delta_c}, g_0 \sqrt{\langle a^\dagger a\rangle}$. Here, $g_0$ is a characteristic zero-photon coupling between the cavity mode and the atomic transitions. Spontaneous emission is not considered in our description, which is a valid approximation if $\abs{\Delta_0}$ is large enough \cite{grimm2000,miller1993}. 

One then obtains an effective Hamiltonian describing cavity field and atomic motion \cite{cormick2013, domokos2001, ritsch2013}: 
\begin{equation}\label{h_efectivo_completo}
  H_{\textrm{eff}} = -\hbar  \Delta_{\textrm {eff}}(x) a^\dagger a
            - i \hbar  \eta \left( a - a^\dagger  \right) + H_{\textrm{ion}}(x),
\end{equation}
\noindent 
where a vector $x$ ($p$) with components $x_j$ ($p_j$) has been defined to simplify the notation.
The operator $\Delta_{\textrm {eff}}(x)$ is defined as
\begin{equation}\label{desintonia_efectiva}
  \Delta_{\text{eff}}(x)=  \Delta_c - \sum_{j=1}^N \frac{g^2(x_j)}{\Delta_0} \equiv \Delta_c - U_0 \sum_{j=1}^N f(x_j)
\end{equation}
\noindent where in the denominator $\Delta_c$ has been neglected in comparison with $\Delta_0$, and $U_0$ is defined as $U_0=g_0^2/\Delta_0$.
The spatial profile of the field in Eq. \eqref{desintonia_efectiva} is chosen to be:
\begin{equation}
  g^2(x_j) = g_0^2 \, f(x_j) \,, \quad  f(x_j)=\cos^2(k x_j)
\end{equation}
\noindent with $k=2\pi/\lambda$, in such a way that the optical potential has a periodicity of $\lambda/2$.
The dispersive coupling in Eq.~\eqref{h_efectivo_completo} can be interpreted as a shift of the cavity frequency determined by the atomic positions, or as a globally deformable optical potential acting on the ions~\cite{ritsch2013}.

For definiteness, we consider only the case of a laser with $\Delta_0>0$ (blue detuning), such that the ions are pushed to the minima of the field intensity.
For strong laser pumping, the ions' equilibrium positions are expected to be close to minima of the optical potential, while for weak pumping the balance between Coulomb repulsion and trap confinement predominates. In a semiclassical treatment, the change from one regime to the other corresponds to a sharp transition when there is a symmetry breaking in the spatial structure of the ions. This is obtained by making the trap minimum coincide with a maximum of optical potential at position $x=0$ \cite{fogarty2015}. In this way, a very small system can be made to mimic the abrupt behavior corresponding to the sliding-pinned transition, which strictly speaking corresponds to an infinite chain.

%Putting all contributions together, the complete system dynamics are described by 
%
%\begin{equation}
%  \dot \rho = -\frac{i}{\hbar} \left[ H, \rho \right] + \mathcal{D}_\kappa (\rho) + \mathcal{D}_\Gamma (\rho) \,.
%\end{equation}
%In the following we will analyze the steady states of such model under a semiclassical approximation.

%%%%%%%%%%%%%%%%%%%%%%%%%%%%%%%%%%%%%%%%%%%%%%%%%%%%%%%%%%%%%%%%%%%%%%%%%%%%%%%%%%%%%%%%%%%%%%%%%%%

\section{Gaussian semiclassical treatment}
\label{sec:semiclassical}

We now review the semiclassical approximation that will be used to analyze the steady states of our system. We refer the reader to \cite{cormick2013}
for more details. Under this treatment the dynamics of the system are described by small quantum fluctuations around classical equilibrium values. One thus expresses:
\begin{equation}\label{semiclasico}
\begin{lcase}
  a &= \overline{a} + \delta a \\
  x &= \overline{x} + \delta x \\
  p &= \overline{p} + \delta p
\end{lcase}
\end{equation}
\noindent where $\overline{a} \equiv \expec{a}$, $\overline{x_j} \equiv \expec{x_j}$, and $\overline{p_j} \equiv \expec{p_j}$, in such a way that each fluctuation operator has zero mean. 
The quantities referring to the ions are grouped in vectors $\overline x$, $\overline p$, $\delta x$ y $\delta p$. 

The approach includes a linearized treatment of fluctuations, truncating the Heisenberg-Langevin equations to first order in these fluctuations. The effect of noise is modeled through the input-output formalism \cite{gardiner1985}. We note that the limitations of this semiclassical treatment have been discussed at length for the case of a single ion in \cite{kahan2023}.

\subsection{Classical equilibrium configurations}

Neglecting input noise operators and setting $\dot a = \dot x_i = \dot p_i = 0$ leads to the following conditions for the classical equilibrium values:
\begin{equation}\label{aeq}
  \overline a = \frac{\eta}{\kappa -i \Delta_{\textrm{eff}}(\overline x)}; \quad   \overline p_i = 0\,. 
\end{equation}
\noindent In turn, the classical solutions for the ion positions are determined minimizing an effective total potential of the form:
\begin{equation}\label{pot_total}
  V_{\textrm{tot}} \equiv V_{\textrm {eff}}(\overline x) + V_{\textrm{ion}} (\overline x)
\end{equation} 

\noindent where $V_{\textrm{ion}}$ contains the trap and Coulomb potentials, whereas $V_{\textrm{eff}}$ is an effective optical potential \cite{maunz2001} defined as
\begin{equation}\label{pot_efectivo}
  V_{\textrm{eff}}(\overline x)= \hbar \frac{\eta^2}{\kappa}\arctan{\left(-\frac{\Delta_{\textrm{eff}}(\overline x)}{\kappa}\right)}\,.
\end{equation}

For weak pumping, minimizing the total effective potential above produces solutions that are similar to those in absence of optical potential. For sufficiently strong pumping, the ions are forced to fall close to optical minima. A relevant parameter in this system is given by the effective cooperativity $C$:
\begin{equation}
    C= \frac{U_0}{\kappa}=
    \frac{g_0^2}{\kappa \Delta_0}\,.
\end{equation}
This parameter determines the strength of the back-action of the ions on the cavity field, i.e. how much the photon number can depend on the ions configuration. Equivalently, it quantifies how far the effective optical potential in Eq.~\eqref{pot_efectivo} is from the standard optical potential in which back-action is negligible, corresponding to $C\ll1$.

\subsection{Fluctuations}

Once the classical equilibrium values have been found, we proceed to find the equations of motion for the fluctuation operators linearizing the Heisenberg-Langevin equations about equilibrium~\cite{cormick2013}. We introduce a normal-mode decomposition of the ion motion without taking into account the presence of cavity field fluctuations. Using the effective potential in Eq. \eqref{pot_efectivo}, the ions equations of motions can be written as 
\begin{equation}
  \ddot \delta x_j = -\frac{1}{m} \sum_i \delta x_i  \partial_{i j} V_{\textrm{tot}}\,.
\end{equation}
\noindent From this linearized equation of motion one finds the normal modes of the ion chain with frequencies $\omega_n$, $n=1,2,3$, which satisfy:
\begin{equation}\label{ec_autovalores_modos_normales}
  K \delta x = m \omega_n^2 \delta x
\end{equation}
\noindent where $K$ is such that $K_{ij}=\partial_{i j} V_{\textrm{tot}}$. Normal mode operators $u_j$ are defined by $u = M^{-1} \delta x$, 
where $M$ is the orthogonal change of basis to the eigenbasis of $K$.
It is important to notice that the normal modes $u_j$ depend on the optical potential through $V_\textrm{eff}$. After the modes have been identified, noise is added according to Eqs.~\eqref{eq:photon losses} and~\eqref{disipador_ion}.

A relevant set of quantities for the study of steady-state entanglement within our semiclassical approximation is given by the couplings $c_n$ of each vibrational mode with the cavity field fluctuations. These are of the form \cite{cormick2013}:
\begin{equation}\label{acoplamientos}
\begin{aligned}
  c_n & = - \sqrt{\frac{\hbar}{2m\omega_n}} \sum_s M_{sn} \nabla_s \Delta_\textrm{eff}(x)\,. \\
\end{aligned}
\end{equation}
\noindent We note that the couplings $c_n$ have dimension of frequency and that by construction, there is no direct coupling between different vibrational modes.

This truncated treatment of the fluctuations can lead, for certain parameter regimes, to unstable behavior. A detailed analysis of stability conditions is carried out in \cite{cormick2013}. Following this work, we restrict our choices of cavity detuning to the range $\Delta_c<0$ to prevent cavity heating of the motion.

\section{Semiclassical treatment of the transition in the three-ion chain}
\label{sec:sp_transition}

In the following we will describe the semiclassical solution for a chain of three ions. Values used in this work (unless stated otherwise) are $\kappa/2\pi=0.2$~MHz, $C=0.5$, $U_0=C \kappa$, while Yb ions are considered and we take $\lambda=369$~nm. The trap frequency is chosen to be $\omega/2\pi\sim0.5$~MHz (see Appendix~\ref{sec:validity} for a discussion of the parameter regime).
The noise on the motional degrees of freedom is chosen to act at a rate $\Gamma_i=10^{-6}\kappa$ and with $N=10$. In the following, when considering frequencies we will often divide by $\kappa$ to obtain dimensionless quantities. This does not necessarily make $\kappa$ the most relevant parameter; it is simply one that is usually fixed in each experimental setup, as opposed to the trap frequency and the pumping strength, and it sets a reference relaxation scale. Similarly, for convenience we will use $x_0=\lambda/4$ as a reference length.

\subsection{Small-scale sliding-pinned transition}

When there is no optical potential, the equilibrium positions of the ions can be analytically determined \cite{james1998}. The central ion is located at the trap center and the interparticle distance is given by:
\begin{equation}
  d_0 =  \left( \frac{5}{16} \frac{q^2}{\pi \epsilon_0 m \omega^2} \right)^{1/3} \!.
  \label{ec:d0}
\end{equation}
In our study, the only variable parameter in this equation will be assumed to be $\omega$, and thus changes in $\omega$ will be equivalent to variations in $d_0$.

In presence of intracavity field, we obtain the classical equilibrium positions of the ions through minimization of the effective potential in Eq. \eqref{pot_total}. For each equilibrium configuration, we will describe the state of the system through the localized semiclassical approximation explained in section \ref{sec:semiclassical}.

When the laser pumping intensity is low, the impact of the optical potential on the equilibrium positions is small, as shown schematically in Fig. \ref{esquema_transicion_FK}-a). In this case, the central ion is still located at the trap center and the configuration has reflection symmetry.
By analogy to the Frenkel-Kontorova model, this configuration is called the ``sliding phase''. We note, however, that the name can be misleading since there is no motional mode with vanishing frequency in this phase, and even in the original FK model the sliding phase is not frictionless \cite{braun2013}. 
More strictly, we will call this parameter regime the ``symmetric phase''.

As the intensity of the laser pump increases, the optical potential becomes more relevant and the particles tend to localize at its minima. This situation is shown in Fig. \ref{esquema_transicion_FK}-b). Since the trap center is aligned with a maximum of the optical potential, optical minima are located at odd multiples of $x_0=\lambda/4$. Thus, in this case, there are two classical solutions that when considered individually no longer have reflection symmetry. We refer to these configurations as ``pinned'' by the optical potential, or belonging to the ``symmetry-broken phase''.

\begin{figure}[t]
\centering
\includegraphics[width=1\columnwidth]{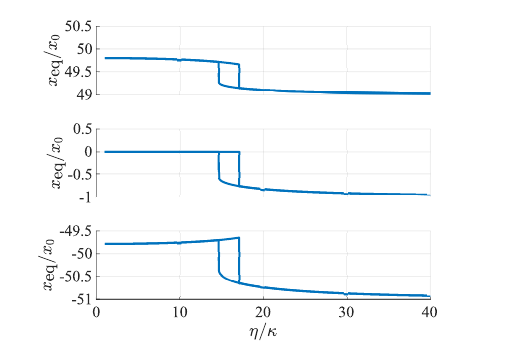}
\caption{Classical equilibrium position of each ion (in units of $x_0=\lambda/4$) as a function of the pumping strength for $C=0.5$, $\Delta_c=0$, and 
$d_0/x_0=49.795$
(see main text for the remaining parameters). For weak pumping, the configuration is symmetric with respect to the trap center; for strong pumping, the ions tend to localize close to minima of the optical potential and thus the symmetry is broken. Both kinds of solutions are found for intermediate laser intensities. For the parameters of this plot, the interparticle distances are almost uniform for all pump strengths.}
\label{transicion_vs_eta}
\end{figure}

For large enough $C$, at intermediate pumping values the semiclassical treatment can predict bistability, i.e. coexistence of different solutions for the ion positions, each one associated with a different equilibrium photon number. An example transition is displayed in Fig. \ref{transicion_vs_eta}, showing the equilibrium positions as a function of the pumping strength and exhibiting bistability for intermediate laser amplitudes. We note that the appearance of classical bistability depends not only on $C$ but also on other parameters such as the cavity detuning $\Delta_c$ \cite{Kahan_2021}. For definiteness, we assume that in the symmetry-broken phase the central ion is always located at negative values of $x$. 

We remark that within the symmetry-broken phase there are two mirror-symmetric solutions for the ion positions, both with equally large mean photon numbers. In the context of this work, when we refer to bistability we speak of a situation where both symmetric and symmetry-broken states (with different photon numbers) are stable steady states for the same parameter values.
From previous work \cite{Kahan_2021,kahan2023}, we know that the validity of the semiclassical Gaussian approximation of sec. \ref{sec:semiclassical} is especially problematic in the classically bistable regime, so from now on we will only analyze the steady states outside the bistability region. According to \cite{Kahan_2021,kahan2023}, the coexistence of two mirror-symmetric solutions does not compromise the validity of the semiclassical approach.

\begin{figure}[htpb!]
\centering
\includegraphics[width=1\columnwidth]{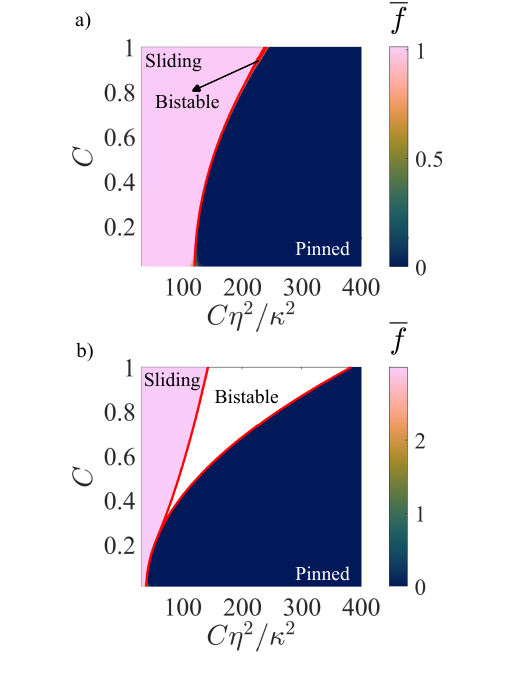}
\caption{Typical phase diagrams, corresponding in this case to $\Delta_c=0$ and
a) $d_0/x_0=51.06$, b) $d_0/x_0=49.98$
 (see main text for the remaining parameters). For comparison with previous articles we refer to ``sliding'' and ``pinned'' phases for symmetric and symmetry-broken regimes respectively. The classically bistable regime is shown between solid red lines. The color code in this graph represents the classical value of $\overline f = \sum_{j=1}^N f(x_j)$, quantifying how far the ions are from being localized at optical minima. \label{diagrama_de_fase}
 }
\end{figure}

The quantity $\overline f  = \sum_{j=1}^N f(x_j)$ indicates how localized the ions are within the optical potential wells, approaching zero as the pumping is increased. In the symmetric phase, $\overline f \geq 1$ as the central ion is located at an optical maximum.
In this way, $\overline f $ exhibits a jump at the transition from a sliding to a pinned phase. We plot this quantity in Fig.~\ref{diagrama_de_fase}, showing typical phase diagrams for this system as a function of the cooperativity $C$ and a representative dimensionless optical potential depth given by the combination $C\eta^2/\kappa^2$ (while $C$ contains the magnitude of the single-photon dispersive coupling, for $\Delta_c=0$ the mean photon number scales roughly as $\eta^2/\kappa^2$). 

We note that the phase diagram strongly depends on $\Delta_c$ and $\omega$. The dependence on $\Delta_c$ is due to the impact of this quantity on the mean photon number, according to Eq.~\eqref{aeq}. On the other hand, the value of the trap frequency determines the equilibrium distances of the ions in absence of optical potential, Eq.~\eqref{ec:d0}, which leads to two different forms of spatial organization in the pinned phase, as will be discussed in the following subsection. This aspect has a strong effect on the character of the transition: as one can observe in Fig.~\ref{diagrama_de_fase}, the critical cooperativity above which the transition becomes discontinuous for the parameters in the figure varies from $C\approx 0.24 $ to $C\approx 0.85$ depending on the ratio $d_0/x_0$.

\subsection{Equilibrium configurations in the pinned phase: uniform and non-uniform chains}

\begin{figure}[htpb!]
\centering
\includegraphics[width=1\columnwidth]{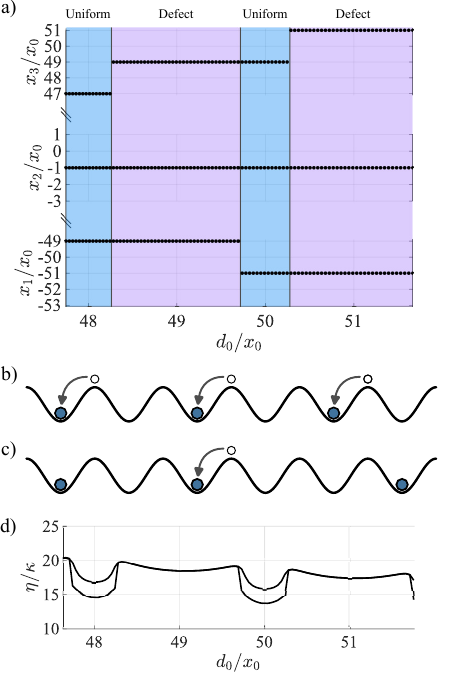}
\caption{a) Equilibrium positions of the three ions (in units of $x_0=\lambda/4$) as a function of $d_0/x_0$ in the deeply pinned regime, with $\eta=400\kappa$, $\Delta_c=0$ (see main text for more parameters). The configuration in the deeply pinned phase can have uniform or non-uniform interparticle distances, as illustrated in subplots b) and c) respectively. Subfigure d) shows the critical pumping strengths as a function of $d_0/x_0$: the upper region corresponds to the pinned phase, the lower region is the symmetric phase, and between the two lines both kinds of solutions coexist.
\label{pos_vs_omega}}
\end{figure}

We now analyze in more detail the possible solutions for the symmetry-broken regime, focusing on the ratio between the ``free'' interparticle distance $d_0$, Eq.~\eqref{ec:d0}, and the potential periodicity $2 x_0=\lambda/2$. The impact of this ratio on the friction experienced by the ion chain was studied experimentally in \cite{bylinskii2015, bylinskii2016observation} for chains of up to 6 ions, in the limit of low cooperativity.

For the specific example in Fig.~\ref{transicion_vs_eta} the ions arrange in a chain with almost uniform interparticle distances. This is true for all values of the pumping strength; even in the pinned phase, the difference between the two distances is of only a few percent of the periodicity of the optical potential. For simplicity, we call this the ``uniform'' parameter regime. For other ratios between the cavity-free interparticle distance $d_0$ and the optical periodicity $2x_0$, the chain can form defects. In this case, in the symmetry-broken phase the ions localize in lattice sites in such a way they are not equidistant. In Fig.~\ref{pos_vs_omega}~a) we display the position of each ion in the deeply pinned phase as the trap frequency is varied, showing the repetitive alternation of uniform and defective regimes.

Two special kinds of cases can be readily analyzed. In the first of these situations, for $\eta\simeq0$ the three ions sit at maxima of the optical potential. Then, in the pinned phase the chain is uniformly displaced with respect to its position in the symmetric phase, as shown in
Fig. \ref{pos_vs_omega}~b).
These cases correspond to the condition
\begin{equation}\label{punto_uniforme}
  d_0 = 2 n x_0, \quad n \in \mathbb{N} 
\end{equation}
and we refer to them as ``matching''. In particular, for the parameter range in Fig.~\ref{pos_vs_omega}, the cases $d_0/x_0=48$ and 50 correspond to $\omega/\kappa\simeq2.70$ and 2.54 respectively. 

The second particular situation occurs when the equilibrium positions of the ions at the ends of the chain in absence of optical potential correspond to optical potential minima. This happens when
\begin{equation}\label{punto_defectos}
  d_0 = x_0 \left( 2n+1 \right), \quad n \in \mathbb{N}
\end{equation}
In this case, increasing the pumping results mostly in a displacement of the central ion, as shown in 
Fig. \ref{pos_vs_omega}~c). 
This ion moves towards a neighboring potential well, while the ions at the ends of the chain remain close to their original positions. This results in a chain with unequal inter-ion distances in the pinned phase, which we name a ``defect''. We refer to these cases as ``maximally mismatched''. For the parameter regime in Fig.~\ref{pos_vs_omega} the condition $d_0/ x_0$ = 49 and 51 results in $\omega/\kappa \simeq 2.62$ and 2.46 respectively.

Both cases commented above correspond to very particular values of $d_0/x_0$. In general, we say that a given value of $d_0/x_0$ corresponds to the uniform regime when the interparticle distances become more uniform as $\eta$ is increased within the pinned phase. Otherwise, we say the chain forms a defect, or corresponds to the non-uniform regime.  We note that the relative size of the regions (i.e., uniform or defective) changes as $d_0/x_0$ is varied. 
We also observe in Fig.~\ref{pos_vs_omega}~d) that for values of $d_0/x_0$ corresponding to the uniform regime, comparatively smaller values of the pumping strength are necessary to reach the pinned phase. This is reasonable since the formation of defects involves overcoming a larger energy barrier due to Coulomb repulsion. 

\begin{figure}[t]
\centering
\includegraphics[width=1\columnwidth]{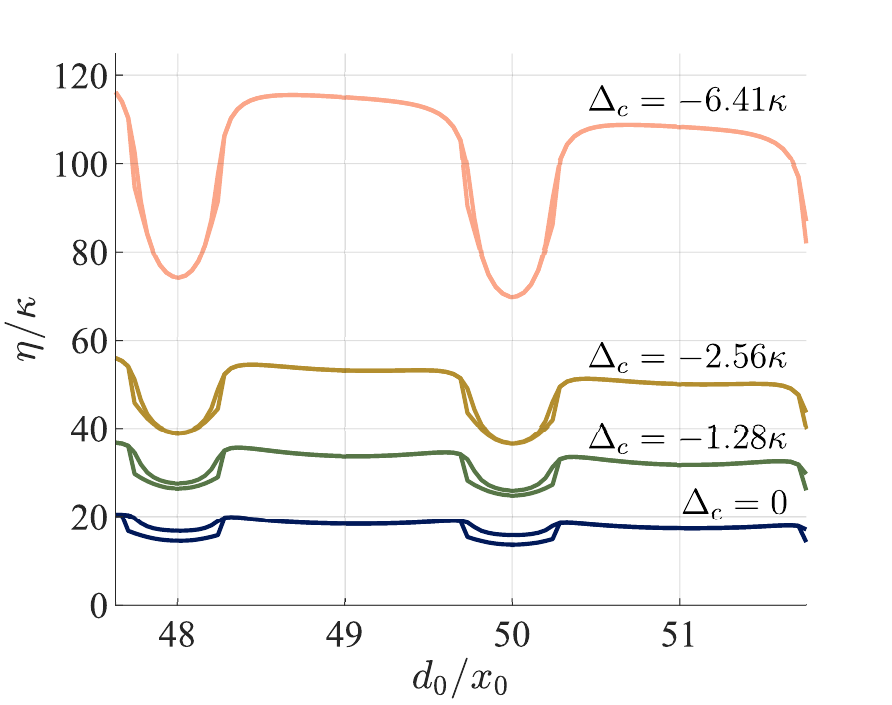}
\caption{Critical pumping strengths as a function of $d_0/x_0$: the upper region corresponds to the pinned phase, the lower region is the symmetric phase, and between the two lines both kinds of solutions coexist. The curves are shown for $C=0.5$ and different values of $\Delta_c$.
(see main text for the remaining parameters).
\label{region_biestabilidad}}
\end{figure}

Finally, one can notice in Fig.~\ref{pos_vs_omega}~d) that for $\Delta_c=0$ and $C=0.5$ the bistable region is only present in the uniform regime. This agrees with the phase diagrams of Fig.~\ref{diagrama_de_fase}. However, for other values of $\Delta_c$ bistability is also absent in the vicinity of the matching conditions. We exemplify this in Fig. \ref{region_biestabilidad}, where we show different transition lines exhibiting bistable regimes. For larger values of $|\Delta_c|$, classical bistability is only found close to boundaries of the uniform region.

\begin{figure*}[htpb!]
\centering
\includegraphics[width=1\textwidth]{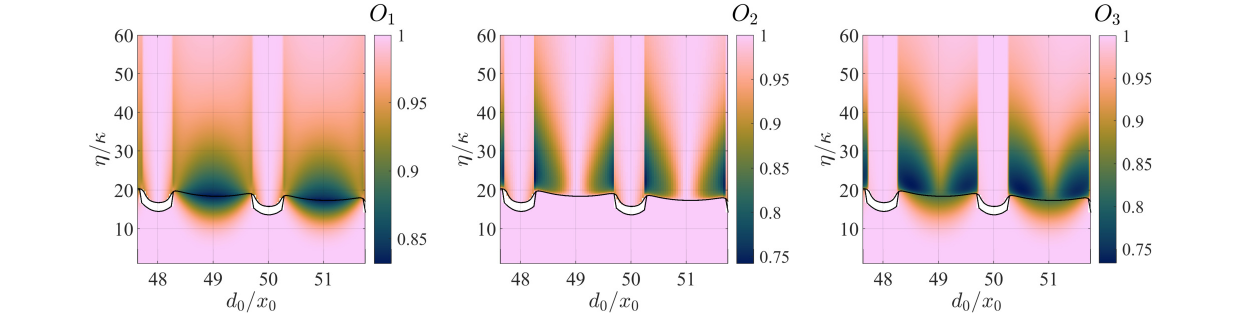}
\caption{Overlaps between the three vibrational modes $u_n(\eta)$ and the reference vibrational modes for vanishing cavity field, $u_n(\eta=0)$. The bistable regime is shown between black solid lines and the overlaps are not computed within this parameter region. See main text for more details.}
\label{overlap_modos_ref_vs_w}
\end{figure*}

\subsection{Vibrational modes in presence of the optical potential}

In absence of optical field, the three normal modes of the ion chain sorted by increasing frequency are \cite{james1998}:
\begin{equation}\label{modos}
\begin{aligned}
  u_1 & = \frac{1}{\sqrt{3}} (\delta x_1 + \delta x_2 + \delta x_3) \\
  u_2 & = \frac{1}{\sqrt{2}} (\delta x_1 - \delta x_3) \\  
  u_3 & = \frac{1}{\sqrt{6}} (\delta x_1 - 2 \delta x_2 + \delta x_3)\,.   
\end{aligned}
\end{equation}
The optical potential causes a deformation of the vibrational modes, even in the sliding phase. 
However, for symmetry reasons, the second motional mode keeps its form as long as the central ion sits at the center of the trap.

To quantify the change in the motional modes, we consider the overlaps $O_n(\eta) = u_n(\eta) \cdot u_n(\eta=0)$, for $n=1,2,3$, between each vibrational mode in the presence of optical potential and the corresponding reference mode vector in Eq.~\eqref{modos}. In Fig. \ref{overlap_modos_ref_vs_w} the color code shows these overlaps as a function of $\eta$ and $d_0/x_0$, for fixed $\Delta_c=0$.
Other choices of $\Delta_c$ shift the transition region in the $\eta$ axis, but the behavior of Fig. \ref{overlap_modos_ref_vs_w} is similar.
We do not observe crossings between vibrational mode frequencies, which would manifest as abrupt jumps in Fig. \ref{overlap_modos_ref_vs_w}.
In particular, for uniform pinned chains the overlaps $O_n$ are always close to 1, so that modes remain very similar to those in absence of cavity field. Fig.~\ref{overlap_modos_ref_vs_w}~b) also shows, as expected, that the overlap for the second mode is exactly 1 in the symmetric configuration.

%%%%%%%%%%%%%%%%%%%%%%%%%%%%%%%%%%%%%%%%%%%%%%%%%%%%%%%%%%%%%%%%%%%%%%%%%%%%%%%%%%%%%%%%
\section{Steady-state entanglement}
\label{sec:entanglement}

In this section we study the entanglement in our system in connection with the chain structure and normal modes, taking into account spatial symmetry and presence of defects in the pinned phase. With this aim, we consider different values of the ratio $d_0/x_0$, and also of the detuning $\Delta_c$. As discussed before, in the symmetry-broken regime the system has mirror-symmetric semiclassical solutions and we focus only on one of them. Within this regime, assuming the validity of the semiclassical approximation, the steady state is generally a statistical mixture of the localized solutions. This is not a problem for the quantification of entanglement since these two solutions have support in different regions of the Hilbert space.

% Indeed, we observe that higher entanglement is often associated with resonances between the effective detuning ($\Delta_\textrm{eff}$) and the normal mode frequencies ($\omega_j^v$).

We quantify bipartite entanglement by the logarithmic negativity $E_N$ \cite{vidal2002, adesso2007}, which has the advantage of being easy to compute for Gaussian states. 
For a Gaussian state with covariance matrix $\sigma$ we have:
\begin{equation}\label{negatividad.logaritmica}
  E_{N} = 
    \begin{lcase}
       - \sum_k &\log (2 \widetilde \nu_k) \qquad \text{for} \; k: \widetilde \nu_k < 1/2\\
      & 0    \qquad \; \; \: \qquad \text{if} \; \; \widetilde \nu_k > 1/2 \quad \forall \; k 
    \end{lcase}
\end{equation}
\noindent where $\lbrace \widetilde \nu_k \rbrace$ is the set of symplectic eigenvalues of $\widetilde \sigma$, 
the covariance matrix of the density matrix after partial transposition with respect to one of the two subsystems~\cite{adesso2007}. We stress that $E_N$ can be computed from $\sigma$, and the experimental estimation of $\sigma$ does not require full quantum state tomography \cite{Fluehmann_2020}.

From $E_{N}$ we calculate the maximum entanglement when varying $\eta$, for each choice of $\Delta_c$ and $d_0$:
\begin{equation}
  \mathcal E (\Delta_c, d_0) \equiv \max_{\eta} E_N (\Delta_c,\eta,d_0)
  \label{eq:E def}
\end{equation}
\noindent where the maximum is found by restricting $\eta$ to values outside the classically bistable regime. In a final subsection, we address the case of multipartite entanglement, by means of the witnesses proposed in  \cite{giedke2001_three_mode, adesso2007}.

\begin{figure*}[htpb!]
\centering
\includegraphics[width=1\textwidth]{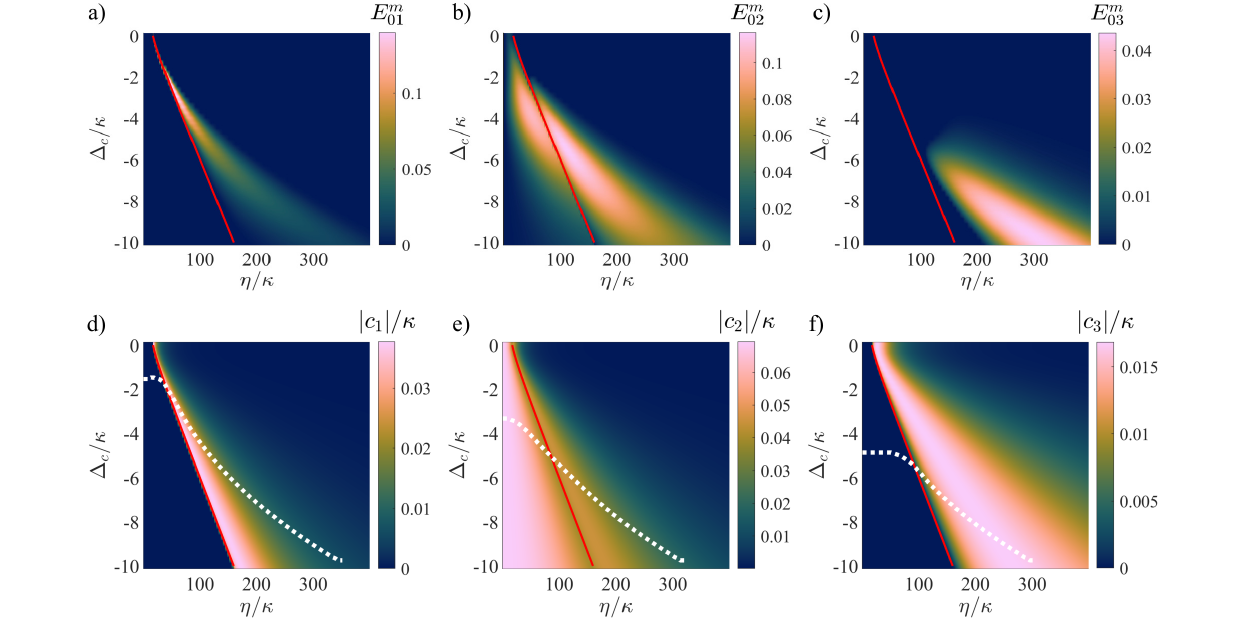}
\caption{a)-c) Entanglement and d)-f) couplings between cavity field fluctuations and each vibrational mode, for different choices of cavity detuning and pumping strength at fixed 
$d_0/x_0=51.44$, corresponding to a non-uniform pinned configuration.
The dotted white lines in d)-f) indicate the resonance between the effective cavity detuning and each of the motional frequencies. In this case, there is no bistability and the red line signals the transition. See main text for the remaining parameters.
}
\label{entrelazamiento_modos_1x1_wok_disp_no_uniforme}
\end{figure*}

The presence of entanglement can strongly depend on the effective temperatures of the different degrees of freedom. We remind the reader that in this work we are including direct noise on the motional modes acting at a low rate, $\Gamma=10^{-6}\kappa$. This noise will have a significant effect on the asymptotic state in the parameter regions in which cavity cooling of the ions is inefficient. For simplicity, we chose the mean excitation number of the noise to be the same for all modes, and for definiteness, we took this number to be $N=10$. The choice of a lower $N$ could lead to entanglement in a larger parameter region. However, it is not the goal of our work to fully characterize the regimes where entanglement appears, but rather to study which configurations are more favorable for it.

\subsection{Bipartite entanglement considering modes as subsystems}

In the following we consider the partition of the system into four subsystems corresponding to the cavity fluctuations and the different vibrational modes. We first analyze bipartite entanglement between two subsystems by tracing out the rest, and then consider bipartitions between one subsystem and the remaining three.

\subsubsection{Entanglement between cavity and each vibrational mode}

We denote by $E_ {ij}^m$ the logarithmic negativity of the reduced state of only 2 subsystems $i,j$ (tracing out the remaining modes), where the indices $i,j=\{0,1,2,3\}$ represent the cavity mode, and the three vibrational modes sorted by increasing frequency, respectively. The superindex $m$ stands for ``mode'', to emphasize the chosen partition, as opposed to the partition in terms of individual ions that will be considered afterwards. First, we will analyze entanglement between the cavity and each vibrational mode, quantified by $E_ {0j}^m$ with $j=1,2,3$.

There are several scenarios where some of the couplings $c_n$ of Eq.~\eqref{acoplamientos} between the cavity field and specific vibrational modes are negligible or zero. This results in insignificant or absent entanglement between those two subsystems. % These situations are summarized in table \ref{latabla}. 
A prominent case is the symmetric or sliding phase, in which $c_1=c_3=0$. Thus, for symmetry reasons in the whole sliding phase the cavity can only become entangled with the second vibrational mode.

A different situation occurs for the second and third modes in the pinned phase without defects. In this case, the couplings $c_2$, $c_3$ are very small. This is due to a combination of two facts: firstly, as was seen in  Fig.~\ref{overlap_modos_ref_vs_w}, in the uniform regime the shapes of the normal modes are almost the same as in absence of the optical potential, Eq.~\eqref{modos}. Secondly, for pinned chains in the uniform regime, the interparticle distances are almost exact multiples of the potential periodicity. Therefore, the contributions to the couplings $c_2, c_3$ in Eq.~\eqref{acoplamientos}  will approximately cancel out. 

\begin{figure*}[htpb!]
\centering
\includegraphics[width=1\textwidth]{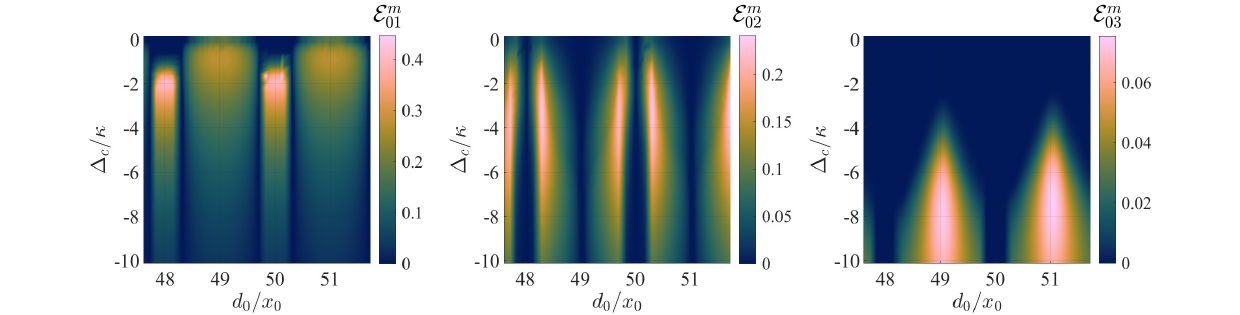}
\caption{Maxima over $\eta$ of the entanglement between cavity field and each vibrational mode, as a function of the cavity detuning $\Delta_c$ and the free equilibrium distance $d_0$ (in units of $\kappa$ and $x_0$ respectively). See Sec.~\ref{sec:sp_transition} for parameters.
\label{entrelazamiento_modos_1x1_omega}}
\end{figure*}

As an example, in Fig.~\ref{entrelazamiento_modos_1x1_wok_disp_no_uniforme} a) to c) we show the negativity $E_N$ between the cavity and each vibrational mode as a function of the cavity detuning and the pump strength for a fixed choice of $d_0$ which corresponds to a non-uniform pinned chain. As expected, the entanglement with the cavity vanishes for the first and third modes in the whole symmetric region. In contrast, for the second mode the entanglement can be non-negligible at both sides of the transition; maxima of negativity with the second mode can be found essentially anywhere on the phase diagram. We note that the locations of the peaks of $E_N$ depend on the particular value of $d_0/x_0$.

The Gaussian semiclassical approximation as described in Sec.~\ref{sec:semiclassical} is more reliable away from the critical points and outside the classically bistable region \cite{Kahan_2021, kahan2023}. As in the example of Fig.~\ref{entrelazamiento_modos_1x1_wok_disp_no_uniforme}, in general one can find significant values of the negativity well separated from the critical line.
This is in contrast with the case of the linear-zigzag transition, where large values of $E_N$ were observed within the classically bistable region and approaching critical points \cite{cormick2013}. 

For comparison, in Figs. \ref{entrelazamiento_modos_1x1_wok_disp_no_uniforme} d) to f) we show the couplings between cavity field fluctuations and each vibrational mode. We see that non-negligible couplings are necessary to obtain entanglement, but entanglement maxima do not generally coincide with maximal couplings.
Indeed, the largest entanglement between the cavity and the $j$-th vibrational mode is generally found close to resonances of the form $\Delta_\textrm{eff}=-\omega_j^v$, indicated with white dotted curves in the lower subplots. This is the reason for the elongated shape of the regions of larger entanglement in Fig.~\ref{entrelazamiento_modos_1x1_wok_disp_no_uniforme} a)-c). Namely, the cavity tends to become more entangled with a given mode when the effective cavity detuning coincides with a so-called ``red sideband'' of the motion. This is a non-trivial fact, since such resonances generally provide a mechanism for cavity cooling \cite{Fogarty_2016}, and complete cooling of the motional mode to its ground state would actually be detrimental for entanglement.

In order to better study the relation between entanglement and chain configuration, in Fig. \ref{entrelazamiento_modos_1x1_omega}  we show, for each vibrational mode, the value of $\mathcal{E}$ as defined in Eq.~\eqref{eq:E def} as a function of the cavity detuning and the ratio $d_0/x_0$. In these plots we observe the same repetitive alternation as in Sec.~\ref{sec:sp_transition}, corresponding to the change between uniform and non-uniform pinned chains.

Entanglement between the cavity field and the first vibrational mode 
$\mathcal E_{01}^m$
in 
Fig. \ref{entrelazamiento_modos_1x1_omega}~a) shows local maxima both in the uniform and non-uniform parameter regime. 
However, uniform chains lead to larger values of logarithmic negativity. As will be noted in section \ref{scaling_cooperativity}, this difference in the maximum values becomes much more drastic for smaller cooperativities.

Entanglement with the second vibrational mode, 
$\mathcal E_{02}^m$
in Fig.~\ref{entrelazamiento_modos_1x1_omega}~b), exhibits maxima predominantly around the boundaries between the uniform and non-uniform regimes.
It is important to note that the Gaussian semiclassical treatment is not very reliable in these parameter regions where different configurations compete.

The entanglement between the cavity fluctuations and the third mode is shown in Fig. \ref{entrelazamiento_modos_1x1_omega}~c). We find negligible entanglement for values of $d_0$ corresponding to the uniform regime. This is because the coupling $c_3$ is exactly zero in the symmetric phase, and it is very small in the pinned phase, as mentioned above.
In the regime of defect formation, the logarithmic negativity can be non-zero, but the maximum values observed are smaller than for the other modes. We note that entanglement with the third mode shows maxima precisely at the points that we referred to as ``maximally mismatched''.

Summing up, the behavior of the entanglement maximum is quite different for each of the vibrational modes: entanglement of the cavity with the first mode is favored by matching conditions, whereas for the third mode the optimal conditions are maximally mismatched, and the second mode becomes more entangled in intermediate cases.

\subsubsection{Entanglement between vibrational modes}

So far we have only discussed bipartite entanglement between the cavity field and each of the vibrational modes. One could equally consider the entanglement between two subsystems corresponding to two different motional modes. However, in our model and under the Gaussian semiclassical approximation we find no entanglement between vibrational modes after tracing out the rest of the system. 

Furthermore, when considering reduced systems that only include two vibrational modes, we also find very small values (of the order of $10^{-4} - 10^{-6}$) of the mutual information \cite{nielsen_chuang}. This means that there are no significant quantum or classical correlations between motional modes.

This might be a consequence of the semiclassical Gaussian treatment.
To determine if this is the case, it would be necessary to extend the analysis of this section to a non-linear treatment of the Heisenberg-Langevin equations of motion in \ref{sec:semiclassical}, or to resort to other kinds of semiclassical approximations as in \cite{kahan2023}. Such a task is beyond the scope of the present paper.

\subsubsection{Bipartitions of the four-mode state}\label{bipartito_cuatro_modos}

We consider now the logarithmic negativities obtained by partitioning the complete system, without taking any partial trace. 
The logarithmic negativity $E_N$ is an entanglement measure for Gaussian states when partitioned into $1\times N$ modes \cite{adesso2007}, so we studied partitions of our system into $1\times 3$ modes. 

The results are shown in Appendix \ref{sec:1x3} for the four possible bipartitions of this form. The plots are not surprising in light of the previous findings: we observe that the logarithmic negativity corresponding to the partition of the system into cavity vs.~ions behaves approximately like the sum of the ones corresponding to the modes taken individually, as in Fig.~\ref{entrelazamiento_modos_1x1_omega}. On the other hand, the entanglement between each motional mode and the rest of the system, including the cavity and the other two vibrational modes, is very similar to the one obtained when tracing out those other two modes.

That is, in terms of vibrational modes the structure of the entanglement in this system is essentially like a star, with the cavity being a central node that is entangled with possibly each of the vibrational modes in an almost additive manner. If the cavity is traced out, there is no entanglement left between normal modes, but each of the modes can be entangled with the cavity.

%--------------

\subsection{Bipartite entanglement considering individual ions}

So far we have analyzed system partitions considering the non-local degrees of freedom given by the collective motional modes. In this section, we analyze the behavior of entanglement when the subsystems are given by the cavity field and the motion of the individual ions. We then use the notation $E_ {ij}^l$, where the superindex stands for ``local''. The subindices $i,j$ run from 0 to 3 as before, but now $1,2,3$ refer to the individual ions labeled from left to right.

\subsubsection{Ion-ion entanglement}

\begin{figure}[t!]
\centering
\includegraphics[width=1\columnwidth]{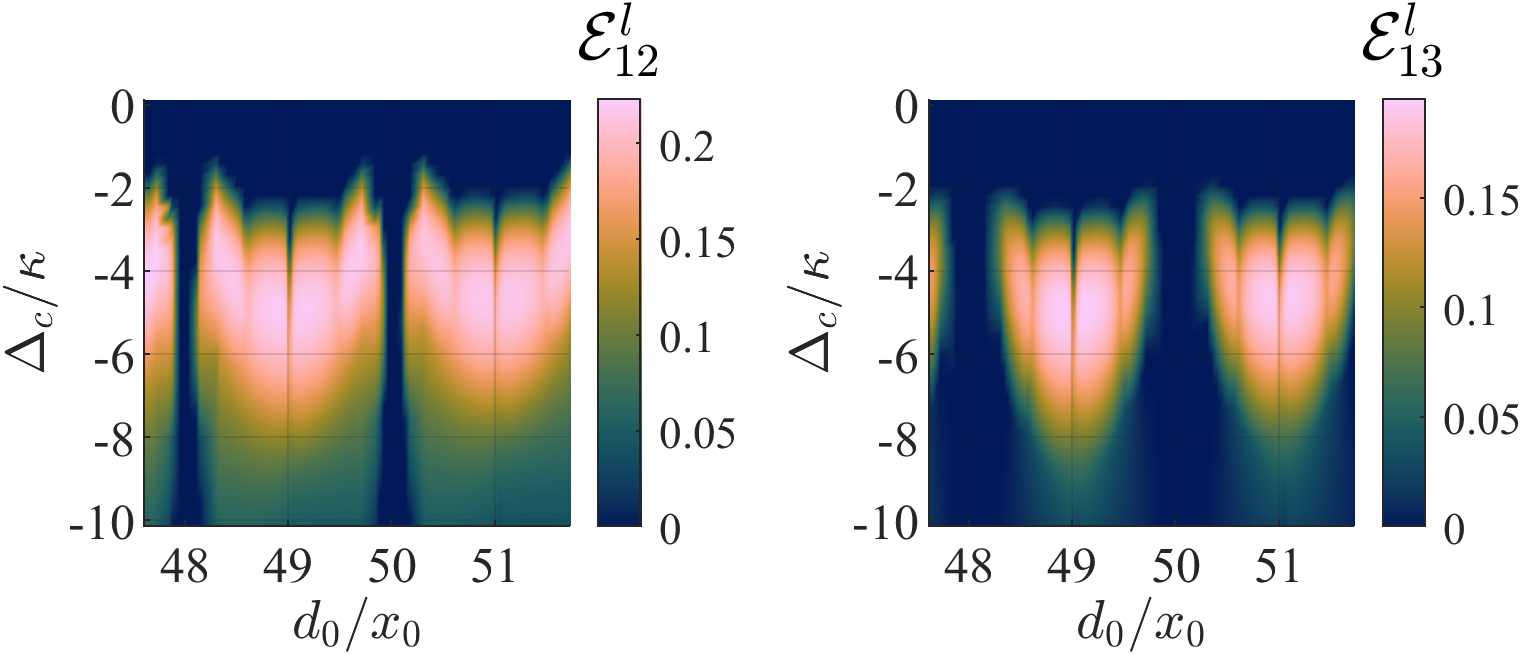}
\caption{
 Maxima over $\eta$ of entanglement between a) consecutive ions, b) ions at the ends of the chain. We note that the maximum values of entanglement between ions 1 and 2 are not exactly the same as between ions 2 and 3, but the difference is too small to be perceived with the scale of this figure.
\label{entrelazamiento_iones_1x1_iones} }
\end{figure}

We first consider the entanglement between two ions, tracing over the cavity and the motional state of the remaining ion. In this case, a much more complex pattern appears, as we show in Fig.~\ref{entrelazamiento_iones_1x1_iones}. Plots a) and b) display the maximum entanglement when varying over $\eta$, for adjacent ions (a) and for the two end ions (b). Here we no longer observe a simple relation between entanglement and chain structure (i.e. uniform or non-uniform regimes). 

Nevertheless, we do find some noticeable patterns: interparticle entanglement vanishes for values of cavity detuning that are close to zero and also in the vicinity of matching conditions. Finally, for our parameter choices inter-ion entanglement is only found in the pinned phase.
This is in contrast with the results in \cite{Retzker_2005}, which focused on inter-ion entanglement in a harmonic trap. However, that work studied ground-state entanglement, whereas for $\eta\to0$ our system is far from the trap ground state due to the inclusion of motional noise. 

We remark that although inter-ion entanglement can be found for relatively wide parameter regions, some of the maxima in Fig.~\ref{entrelazamiento_iones_1x1_iones} are found close to the transition line. Leaving out a region around this line can lead to smaller maximum values, still of the same order of magnitude.

\begin{figure*}[htpb!]
\centering
\includegraphics[width=1\textwidth]{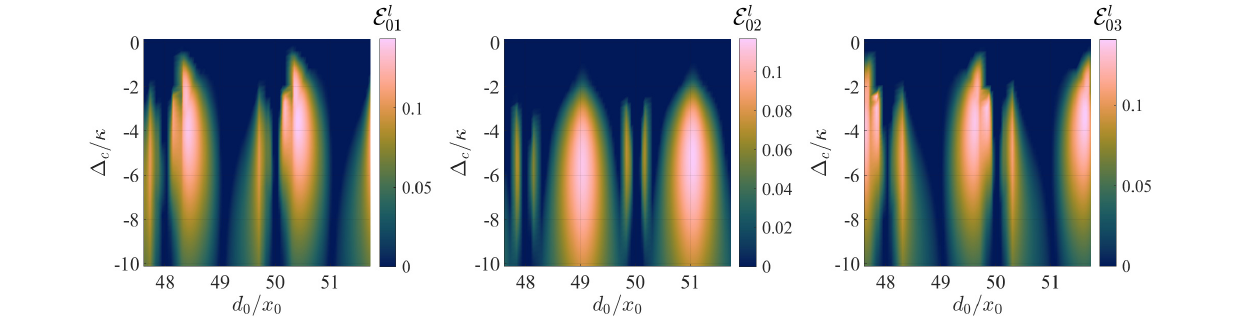}
\caption{
Maxima over $\eta$ of entanglement between cavity fluctuations and individual ions. See Sec.~\ref{sec:sp_transition} for parameters.
\label{entrelazamiento_iones_1x1_omega}}
\end{figure*}

\subsubsection{Entanglement between cavity fluctuations and ions}

We now move on to consider the entanglement between cavity fluctuations and individual ions, tracing over the other two motional degrees of freedom. In the first place, we note that these logarithmic negativities are numerically verified to be zero in the symmetric configuration for all the parameters explored. Entanglement between cavity and individual ions is only found to be non-negligible in the pinned phase.

We then consider the maximum $\mathcal{E}$ of the logarithmic negativity over $\eta$ for each of the bipartitions containing the cavity and one of the individual ions. The results are displayed in Fig.~\ref{entrelazamiento_iones_1x1_omega}. First, we notice that none of the ions is entangled with the cavity for the matching conditions of Eq.~\eqref{punto_uniforme} or for $\Delta_c\simeq 0$. Subplot b) shows entanglement with the central ion for a broad range of parameters, especially within the non-uniform regime. Entanglement with the end ions is generally found to be largest in regions that are neither close to matching nor close to maximally mismatched. 

We note that due to our arbitrary choice of symmetry breaking there is an asymmetry when comparing Figs.~\ref{entrelazamiento_iones_1x1_omega}~a) and c), corresponding to the left and right ends of the chain respectively. The results indicate that the cavity becomes more entangled with the end ion that is displaced from its original position in the same direction as the central ion.

\subsection{Variations with cooperativity}\label{scaling_cooperativity}

Previous work \cite{cormick2013} has observed, for cooperativities of the order of 1, that larger values of $C$ lead to larger entanglement. Here, we study the behavior of our system for a smaller cooperativity of $C=0.1$ and comment on two observations that go beyond that simple rule.

Firstly, while this smaller cooperativity typically leads to lower entanglement, the qualitative behavior of bipartite entanglement remains unchanged, except for the case of the entanglement of the cavity with the lowest-frequency mode. As was mentioned above, for this case we observe that for lower cooperativities the achievable values of entanglement are much smaller in the non-uniform structure than in the uniform regime. 

Secondly, entanglement between neighboring ions remains essentially unaffected by this change in cooperativity. This suggests that entanglement across these bipartitions mostly depends on the chain structure and not on the presence of back-action. Indeed, it is known that at low temperatures different ions can be entangled due to Coulomb interaction \cite{Retzker_2005}.

\subsection{Multipartite entanglement}

\subsubsection{Tripartite entanglement}

In this section, we analyze the presence of tripartite entanglement after tracing over one of the subsystems. 
There are simple forms of quantifying multipartite entanglement in CV systems whenever the states are either pure \cite{adesso2006}, or bisymmetric \cite{serafini2005} (states invariant under exchange of two modes). The state of our system is bisymmetric with respect to some partitions in some parameter regimes. In particular, in the sliding phase the system is invariant under the exchange of the ions at the sides. However, this is not the general case. 

\begin{figure}[htpb!]
\centering
\includegraphics[width=1\columnwidth]{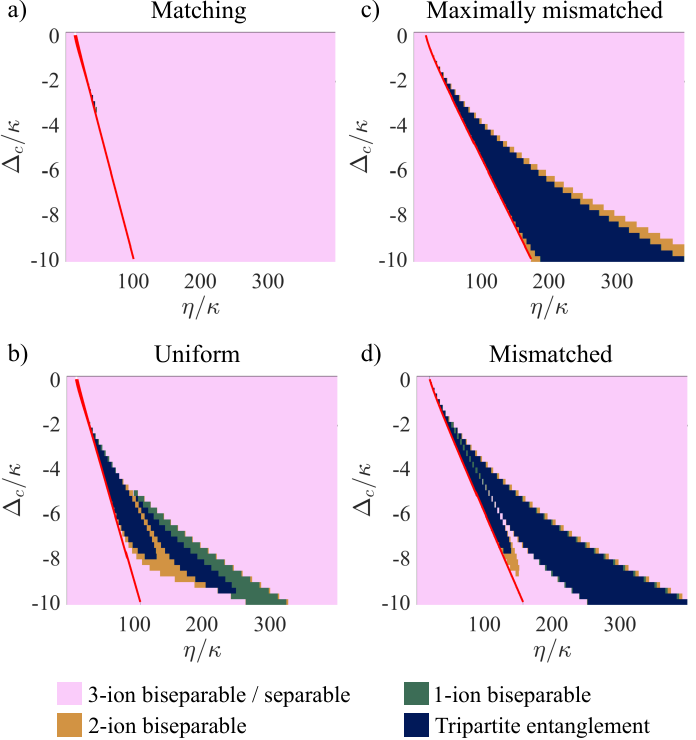}
\caption{Classification of the reduced state of the motional degrees of freedom of the individual ions in terms of their presence of genuine multipartite entanglement. The bistable regime is shown between solid red lines. 
Values of $d_0/x_0$ are a) 49.99, b) 48.99, c) 49.86, d) 48.28. See Sec.~\ref{sec:sp_transition} for the remaining parameters.
\label{fig_tripartito}}
\end{figure}

We then restrict only to detect the presence of tripartite entanglement without quantifying it, through the use of the criteria detailed in \cite{giedke2001_three_mode, adesso2007}.
This method does not rely on solving an optimization problem, in contrast to other methods for detecting multipartite entanglement \cite{shchukin2015}.
Following this criterion, 
Gaussian states are categorized according to the number of bipartitions that make them separable. 
If no bipartition makes them separable, then the state has genuine tripartite entanglement \cite{giedke2001_three_mode}. 

We show this classification in Fig. \ref{fig_tripartito} for the reduced state of the ions, tracing over the cavity, using different colors for the different cases: separable, 2-ion biseparable, 1-ion biseparable, and tripartite entangled. Each of the subplots corresponds to a different choice of $d_0$. 

We see that for the parameters we chose the state is always 3-separable (separable respect to all bipartitions) in the symmetric phase. We stress once more that this is a consequence of our choice of noise, and that entanglement between different ions can be observed for lower temperatures \cite{Retzker_2005}. Within the pinned phase there is a significant region where the state shows genuine tripartite entanglement. The area of this region grows as the system approaches the maximally mismatched condition \eqref{punto_defectos}, and vanishes completely whenever the matching condition of Eq. \eqref{punto_uniforme} is satisfied.

We also observe, in Figs.~\ref{fig_tripartito}~b) and d), that the region of tripartite entanglement is ``cut'' by oblique narrow regions of 2-ion and 1-ion separability. These coincide with curves where the couplings to certain vibrational modes vanish or become extremely small, leading to higher effective temperatures.

By tracing out other subsystems we can consider other forms of tripartite entanglement. There are a few general points to remark. Firstly, upon tracing over any mode or ion, there is no tripartite entanglement whenever the system is close to the maximally matched condition.
Secondly, there is generally no genuine tripartite entanglement in the sliding phase, with just one exception: when tracing out the central ion, there is tripartite entanglement in both the sliding and pinned phases.
Finally, we observe that there is no tripartite entanglement when considering the reduced state of the three vibrational modes.

\subsubsection{Four-partite entanglement}

Having verified the presence of genuine tripartite entanglement within several regimes, we use the multipartite entanglement criteria for CV systems detailed in \cite{vanLoock2003} to verify whether there is four-partite entanglement. In four-mode states, this criterion consists of evaluating the following inequalities:

\begin{equation}\label{vL_ineq}
\begin{aligned}
\textrm{I.} & \langle[\Delta\left(\hat{x}_1-\hat{x}_2\right)]^2\rangle+\langle\left[\Delta\left(\hat{p}_1+\hat{p}_2+g_3 \hat{p}_3+g_4 \hat{p}_4\right)\right]^2\rangle \geq 2 \\
\textrm{II.} & \langle\left[\Delta\left(\hat{x}_2-\hat{x}_3\right)\right]^2\rangle+\langle\left[\Delta\left(g_1 \hat{p}_1+\hat{p}_2+\hat{p}_3+g_4 \hat{p}_4\right)\right]^2\rangle \geq 2 \\
\textrm{III.} & \langle\left[\Delta\left(\hat{x}_1-\hat{x}_3\right)\right]^2\rangle+\langle\left[\Delta\left(\hat{p}_1+g_2 \hat{p}_2+\hat{p}_3+g_4 \hat{p}_4\right)\right]^2\rangle \geq 2 \\
\textrm{IV.} & \langle\left[\Delta\left(\hat{x}_3-\hat{x}_4\right)\right]^2\rangle+\langle\left[\Delta\left(g_1 \hat{p}_1+g_2 \hat{p}_2+\hat{p}_3+\hat{p}_4\right)\right]^2\rangle \geq 2 \\
\textrm{V.} & \langle\left[\Delta\left(\hat{x}_2-\hat{x}_4\right)\right]^2\rangle+\langle\left[\Delta\left(g_1 \hat{p}_1+\hat{p}_2+g_3 \hat{p}_3+\hat{p}_4\right)\right]^2\rangle \geq 2 \\
\textrm{VI.} & \langle\left[\Delta\left(\hat{x}_1-\hat{x}_4\right)\right]^2\rangle+\langle\left[\Delta\left(\hat{p}_1+g_2 \hat{p}_2+g_3 \hat{p}_3+\hat{p}_4\right)\right]^2\rangle \geq 2. \\
\end{aligned}
\end{equation}

\noindent where the $g_i$ are arbitrary real constants, and $\Delta^2 (\hat q)$ represents the variance of an observable $\hat q$. The violation of each inequality in \eqref{vL_ineq} implies different entanglement structures for the state, as explained in \cite{vanLoock2003}.

We first consider local degrees of freedom, and evaluate the inequalities above for several choices of $g_i$ which for simplicity satisfy $g_i=g \; \forall i$. For matching conditions, all inequalities are satisfied, and no conclusion can be drawn about the inseparability of the state. Sufficiently far from the matching points, inequalities II and IV can be violated in small regions contained within the regimes where tripartite entanglement is present in Fig.~\ref{fig_tripartito}, in the pinned phase and both in uniform and non-uniform cases. 

\begin{figure}[htpb!]
\centering
\includegraphics[width=1\columnwidth]{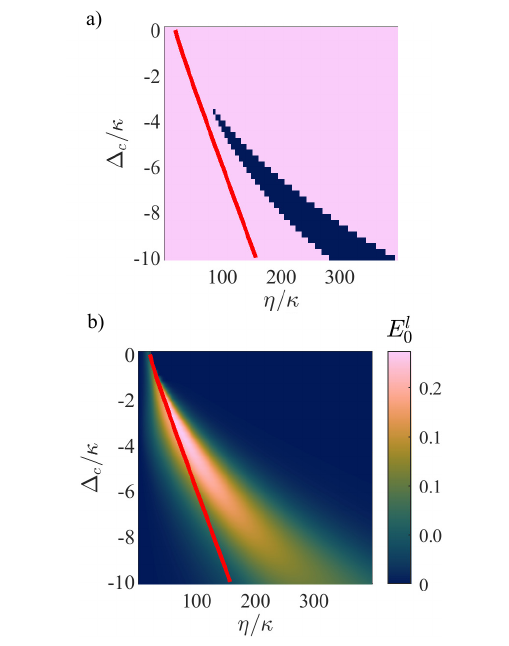}
\caption{a) Classification of the state of the complete system for the motional degrees of freedom of the individual ions, showing in blue the region where inequalities II and IV in Eq.~\eqref{vL_ineq} are simultaneously violated, for $g_i=0.1$. b) Entanglement between the cavity and the ion chain, to complement the information from the inequalities \eqref{vL_ineq}. Parameter values corresponding to the blue region in a) and non-zero entanglement in b) are four-partite entangled. The bistable regime is shown between solid red lines, and $d_0/x_0=48.28$. See Sec.~\ref{sec:sp_transition} for the remaining parameters.
\label{four_mode_ent_sep}}
\end{figure}

In Fig.~\ref{four_mode_ent_sep}~a) we show in blue the region where inequalities II and IV are simultaneously violated for the choice $g=0.1$. This value leads to comparatively large regions where multimode entanglement is detectable with this criterion. The simultaneous violation of II and IV implies that the state either has four-partite entanglement or can be written as \cite{vanLoock2003}
\begin{equation}
    \rho = \sum_i a_i \rho_{i,\textrm{ions}} \otimes \rho_{i,\textrm{cav}}
    \label{eq: separable cavity}
\end{equation}
exhibiting tripartite entanglement in the reduced state of the ions. At the maximally mismatched point we observe that inequality V is violated as well, but as explained in \cite{vanLoock2003}, this does not provide any additional information. 

We can complement the information extracted from these inequalities with the value of the negativity between the cavity and the rest of the system, as in Sec. \ref{bipartito_cuatro_modos}. This is displayed in Fig. \ref{four_mode_ent_sep}-b). We observe non-vanishing negativity $E_0^l$ in regions where inequalities II and IV are not satisfied. When this happens, we can exclude the possibility of a state of the form of Eq.~\eqref{eq: separable cavity}, thus concluding that the state must be four-partite entangled.

\begin{figure}[htpb!]
\centering
\includegraphics[width=1\columnwidth]{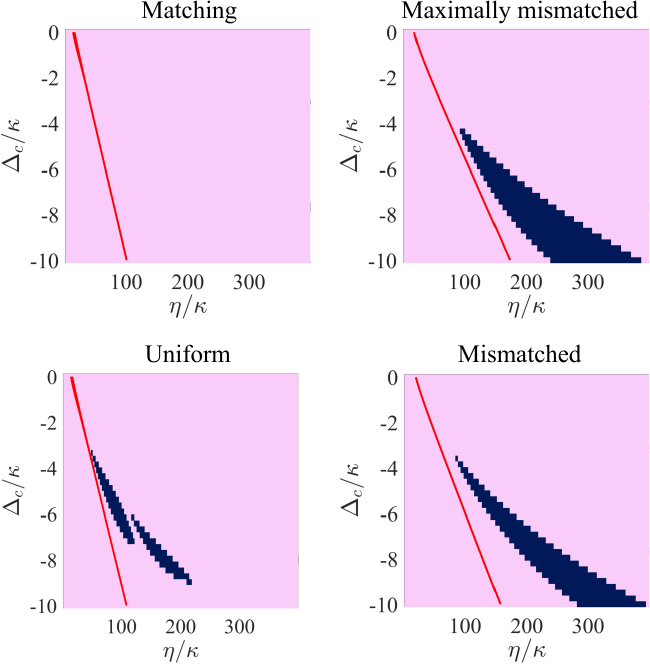}
\caption{Classification of the state in terms of the presence of genuine four-partite entanglement, indicated in blue, with subsystems taken as cavity and individual ions. The possibility of four-partite entanglement in the pink region is not discarded. The diagrams result from the combination of Eq.~\eqref{vL_ineq} evaluated at $g_i=0.1$ and the entanglement between cavity and ion chain. The bistable regime (or transition point for the mismatched cases) is shown between solid red lines.
Values of $d_0/x_0$ are a) 49.99, b) 48.99, c) 49.86, d) 48.28. See Sec.~\ref{sec:sp_transition} for the remaining parameters.
}
\label{cuatripartito}
\end{figure}

We extend this argument in Fig.~\ref{cuatripartito} for different ratios of $d_0/x_0$ (the same values as in Fig. \ref{fig_tripartito}). We show in blue the regimes where both inequalities II and IV are violated while there is non-vanishing entanglement between the cavity and the rest of the system at the same time. This shows that, considering subsystems given by the cavity and the motion of the individual ions, four-partite entanglement can be found in the pinned phase whenever the system is far enough from the matching condition. We also note that the regions with four-partite entanglement are contained within those where the ion chain exhibits tripartite entanglement (see Fig.~\ref{fig_tripartito}).

In terms of the normal mode coordinates, we could not find any pairs of inequalities in \eqref{vL_ineq} that are simultaneously violated for any of our 16 choices of $g_i$ between $-1$ and 1.
%{\color{blue} $g_i= -1,\, -0.5,\, -0.1,\, -0.05,\, -0.01,\, -0.005,\,-0.001,\, 0,\, 0.001,\, 0.005,\, 0.01,\, 0.05,\, 0.1,\, 0.2,\, 0.5,\, 1 $} 
This set of criteria is therefore inconclusive about the presence of four-mode entanglement.

\section{Conclusions}
\label{sec:conclusions}

We studied entanglement across a structural transition in a one-dimensional chain of three ions dispersively coupled to an optical potential. We identified the equilibrium chain configurations as sliding or symmetric and pinned or symmetry-broken. In the latter case we further observed the presence of regimes leading to uniform and non-uniform chains. 

We used a semiclassical approximation of localized Gaussian states, 
which is expected to be valid outside the region where symmetric and symmetry-broken solutions coexist, and far enough from the transition points. In contrast with the case of the linear-zigzag transition studied in  \cite{cormick2013}, we found that it is possible to obtain non-negligible entanglement well outside those problematic parameter ranges.

Our results show strong correlations between the equilibrium chain structure and the entanglement across system bipartitions. Indeed, different chain configurations tend to lead to different kinds of entanglement. For example, we observe larger amounts of entanglement between the cavity field and the lowest motional mode for pinned uniform chains, but larger entanglement between the cavity and the central ion for pinned non-uniform chains.

We also analyzed the presence of tripartite entanglement within all reduced three-subsystem states, as well as four-partite entanglement. We concluded that the formation of defects in the structure leads to larger parameter regimes of multipartite entanglement. On the contrary, when the equilibrium distances between ions in absence of optical potential match the distance between optical wells, the criterion used fails to detect multipartite entanglement.

Finally, we noticed that entanglement is favored by the presence of ``red-sideband resonances'' between the cavity detuning and the motional frequencies.
Indeed, the vicinity to these resonances enhances cavity cooling of the vibrations, and therefore the formation of asymptotic states with low entropy. Non-trivially, these resonances lead to higher entanglement not only when the chain subsystems are chosen to be the individual ions but also when they correspond to collective modes.

Because of the formalism applied, in this work we focused on entanglement outside the parameter regions with semiclassical bistability. It is important to stress that we could find four-partite entanglement far from any transition points. However, previous work \cite{cormick2013, Kahan_2021} suggests that the build-up of entanglement could be favored by the coexistence of different classical configurations. More sophisticated numerical techniques are necessary to explore this possibility.

\section{Acknowledgments}

The authors acknowledge funding from Grant PICT 2020-SERIEA-00959 from ANPCyT (Argentina).

\appendix

\section{Bipartitions of the four-mode state}
\label{sec:1x3}

We now consider the logarithmic negativities obtained by partitioning the complete system, i.e. without taking any partial trace, into $1\times 3$ modes. 
That is, here we quantify the entanglement between subsystem $A$ and the remaining 3 modes, and we denote it by 
$E_{j}^m$, with $j=\{0,1,2,3\}$ representing the subsystem that is traced out. In analogy with the notation already used, we denote its maximum over $\eta$ by $\mathcal E_{j}^m$.

\begin{figure}[htbp!]
\centering
\includegraphics[width=1\columnwidth]{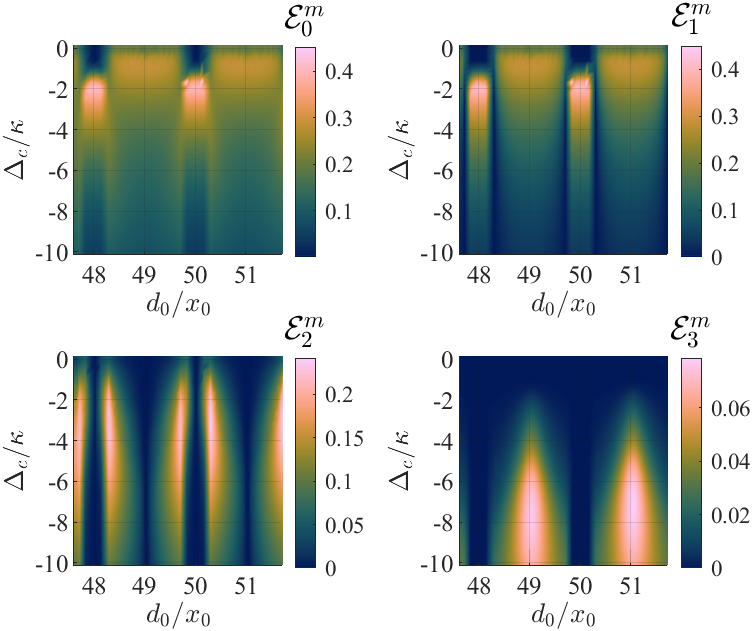}
\caption{Maxima over $\eta$ of entanglement between any single mode and the rest of the system.}
\label{entrelazamiento_modos_1x3}
\end{figure}

The results are shown in the Fig. \ref{entrelazamiento_modos_1x3} for the four possible bipartitions of this form. In subplot a), which corresponds to the partition cavity -- collection of motional modes, we observe that the entanglement 
$\mathcal E_0^m$ behaves approximately like the combination of the negativities of the two-mode bipartitions including the cavity field (see Fig. \ref{entrelazamiento_modos_1x1_omega}). 
On the other hand, the negativities 
$\mathcal E_j^m$ with $j=\{1,2,3\}$
shown in Fig. \ref{entrelazamiento_modos_1x3}~b) to d) give quantitatively similar results to the 
same negativities individually.

\section{Parameter regime and validity of the treatment applied}
\label{sec:validity}

In this Appendix we discuss the validity of the localized semiclassical approximation used. In general, the ions need not be localized around only one classical equilibrium, and the configuration in which they have maximal probability to be found may not coincide with the global minimum of the effective potential in Eq.~\eqref{pot_total}, see \cite{kahan2023}. Furthermore, even if this is the case, the kinetic energy corresponding to the vibrational modes could cause diffusion towards adjacent local minima. In order to determine a parameter regime in which this has low probability, we estimate the energy cost of shifting the equilibrium configuration to adjacent minima.

This estimation is based on the semiclassical effective potential, Eq.~\eqref{pot_total}. We note that this potential is generally only a means to find all possible semiclassical equilibrium configurations and is not a true energy. However, in the limit of small cooperativities it does correspond to the standard optical potential. Furthermore, in~\cite{kahan2023} it was confirmed that when the difference in effective potential between competing configurations is large enough, then the most likely configuration tends to coincide with the global minimum.

\begin{figure}[htpb!]
\centering
\includegraphics[width=1\columnwidth]{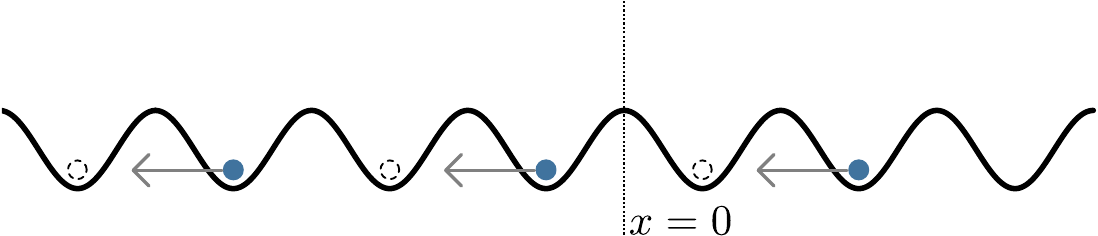}
\caption{The configuration corresponding to the global minimum of the effective potential is shown in blue, in the pinned phase with uniform displacement. The energy cost of uniformly moving the chain to the configuration shown with dashed lines as schematized in this figure is considered. The dotted vertical line shows the center of the harmonic trap. The trap potential is not illustrated in this schematic representation.}
\label{dif_energias}
\end{figure}

We want our choice of parameters to be such that the energy cost of uniformly displacing the chain in the pinned phase over a period of the optical potential ($\lambda/2$) is larger than the vibrational energy. That is, we consider the energy difference between the two configurations shown in Fig.~\ref{dif_energias} and compare it with the energy in the vibrational modes. When the former is larger, we assume that we can neglect the possibility of statistical superpositions involving several optical wells for each of the ions.

This criterion means that our treatment is not valid for low trap frequencies. This is, our values of $d_0/x_0$ around 50 are chosen to avoid diffusion between different local minima of the potential \cite{Katori_1997} while keeping the frequency $\omega$ in a realistic range. We note however that if one applies the same treatment for smaller values of $\omega$, leading to larger values of $d_0/x_0$, the results do not change drastically but they become less reliable.

% ------------ Bibliografia --------------
% Para biblatex
% \printbibliography

% Para natbib
% \bibliography{biblio.bib}


% ----------------------------------------

\end{document}